\documentclass[showpacs,amsmath,superscriptaddress,
reprint,aps]{revtex4-1}

\usepackage{hyperref}
\usepackage{graphicx}

\begin{document}

\title
{Room-temperature single photon emission from micron-long\\air-suspended carbon nanotubes}
\author{A.~Ishii}
\affiliation{Nanoscale Quantum Photonics Laboratory, RIKEN, Saitama 351-0198, Japan}
\affiliation{Quantum Optoelectronics Research Team, RIKEN Center for Advanced Photonics, Saitama 351-0198, Japan}
\author{T.~Uda}
\affiliation{Nanoscale Quantum Photonics Laboratory, RIKEN, Saitama 351-0198, Japan}
\affiliation{Department of Applied Physics, The University of Tokyo, Tokyo 113-8656, Japan}
\author{Y.~K.~Kato}
\email[Corresponding author. ]{yuichiro.kato@riken.jp}
\affiliation{Nanoscale Quantum Photonics Laboratory, RIKEN, Saitama 351-0198, Japan}
\affiliation{Quantum Optoelectronics Research Team, RIKEN Center for Advanced Photonics, Saitama 351-0198, Japan}

\begin{abstract}
Statistics of photons emitted by mobile excitons in individual carbon nanotubes are investigated. Photoluminescence spectroscopy is used to identify the chiralities and suspended lengths of air-suspended nanotubes, and photon correlation measurements are performed at room temperature on telecommunication-wavelength nanotube emission with a Hanbury-Brown-Twiss setup. We obtain zero-delay second-order correlation $g^{(2)}(0)$ less than 0.5, indicating single photon generation. Excitation power dependence of the photon antibunching characteristics is examined for nanotubes with various chiralities and suspended lengths, where we find that the minimum value of $g^{(2)}(0)$ is obtained at the lowest power. The influence of exciton diffusion and end quenching is studied by Monte Carlo simulations, and we derive an analytical expression for the minimum value of $g^{(2)}(0)$. Our results indicate that mobile excitons in micron-long nanotubes can in principle produce high-purity single photons, leading to new design strategies for quantum photon sources.
\end{abstract}

\maketitle

\section{Introduction}
Single photon emitters are a fundamental element for many quantum information processing and communication technologies \cite{OBrien:2009}, and the quest for ideal single photon sources have continued since the initial report of photon antibunching in individual atoms \cite{kimble:1977}. Extensive range of systems have been investigated, including ions \cite{diedrich:1987}, molecules \cite{basche:1992, lounis:2000}, and nanocrystals \cite{michler:2000nature, choi:2014}. In particular, color centers in diamond and SiC offer room-temperature single photon emission since their electronic states have sufficient separation from the valence and conduction bands \cite{Kurtsiefer:2000,castelletto:2014}. Unfortunately, the emission wavelengths of these color centers are in the visible range where attenuation through optical fibers is large. In comparison, semiconductor quantum dots can be designed to generate single photons in the telecommunication band \cite{Benyoucef:2013}, and long-distance quantum key distribution experiments have been demonstrated \cite{Takemoto:2010}. The operation of these quantum dots, however, is limited to cryogenic temperatures. Recent exploration of single photon emission capability has expanded to other materials such as atomically thin materials \cite{He:2015,Tran:2016} and perovskites \cite{park:2015}, but it is still difficult to simultaneously achieve both room-temperature operation and telecommunication-wavelength emission.

In this regard, carbon nanotubes have a potential for becoming an ideal single photon source as their excitonic states are stable even at room temperature \cite{Wang:2005, Maultzsch:2005} and their emission wavelengths cover a broad spectrum including the telecommunication wavelength \cite{Weisman:2003, Ishii:2015}. Localized excitons at low temperatures exhibit single photon emission \cite{Hogele:2008}, whereas air-suspended nanotubes show narrower linewidth and more stable emission \cite{Hofmann:2013}. Optical cavity configurations can be utilized to achieve enhancement of the emission rate \cite{Newman:2012} and tunability of the operation wavelength \cite{Jeantet:2016}, while antibunching has been observed from an electrically-driven nanotube emitter integrated into a photonic circuit \cite{Khasminskaya:2016}. In recent years, room-temperature operation has been demonstrated by taking advantage of exciton trapping sites in oxygen-doped nanotubes \cite{Ma:2015NatNano} and air-suspended nanotubes \cite{Endo:2015}.

All of these efforts, although spanning various classes of materials, are similar in the sense that they utilize individual localized states. In contrast, if mobile excitons can be used for single photon generation, it may lead to a new class of devices with qualitatively different designs and functionalities. Without the necessity for confinement potentials, there would no longer be the need to localize the states using cryogenic environment. If the emitter material hosting the mobile excitons are micron-scale, integration into electronic devices or photonic structures would become straightforward. It may also become possible to employ far-field optical techniques to perform spatiotemporal manipulation of photon emission.

One possible approach to creating single photons from mobile excitons is to utilize efficient exciton-exciton annihilation (EEA) in carbon nanotubes. In this relaxation mechanism, an exciton can recombine nonradiatively by giving its energy to another exciton via Auger-like process \cite{Wang:2004prb, Wang:2006}. It is known that EEA occurs very effectively in nanotubes due to the uniqueness of one-dimensional diffusion, where excitons traverse in a compact manner \cite{Gennes:1982, Ishii:2015}. When multiple excitons exist, EEA causes rapid recombination until there is only one exciton remaining. As EEA does not occur for a single exciton, the last surviving exciton can recombine to emit a single photon. It is not intuitively obvious whether such an emission mechanism allows for high-performance single photon generation, but limitations on the attainable degree of antibunching has been suggested \cite{Ma:2015PRL}.

Here we investigate photon statistics in micron-scale air-suspended carbon nanotubes, and perform theoretical analysis to show that diffusion-driven annihilation of mobile excitons can produce high-purity single photons. Clean, as-grown nanotubes are characterized by photoluminescence (PL) spectroscopy, and photon correlation measurements are done on telecommunication-wavelength nanotube emission at room temperature. The normalized second-order correlation function $g^{(2)}(\tau)$ exhibits clear photon antibunching, where the coincidence is reduced at a delay time $\tau=0$. From the excitation power dependence, we find that smaller $g^{(2)}(0)$ is obtained at lower powers. The values of $g^{(2)}(0)$ seem to be different for nanotubes with different chiralities and suspended lengths, and Monte Carlo simulations are performed to elucidate the effects of exciton diffusion length and nanotube length. An analytical expression for the minimum value of $g^{(2)}(0)$ is derived, which shows that single photon emission performance can be improved considerably.

\section{Photoluminescence spectroscopy and photon correlation measurements}
Our air-suspended carbon nanotubes are grown over trenches on bare Si substrates \cite{Ishii:2015}. We perform electron beam lithography and dry etching to form trenches with widths ranging from 0.2 to 6.8~$\mu$m. Another electron beam lithography step defines catalyst areas near the trenches, and Fe/silica dissolved in ethanol is spin-coated and lifted off. Single-walled carbon nanotubes are synthesized by alcohol chemical vapor deposition at 800$^\circ$C with a growth time of 1 min.

\begin{figure}
\includegraphics{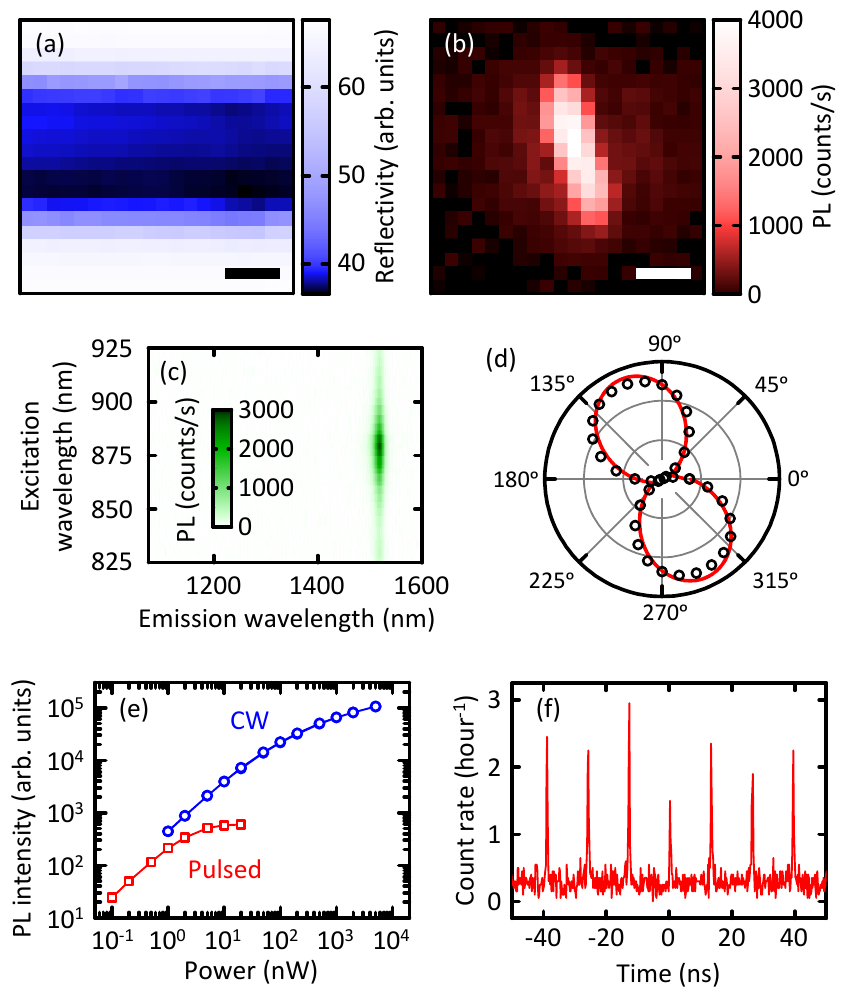}
\caption{
\label{Fig1} (a) and (b) Reflectivity and PL images, respectively. An excitation power of 2~$\mu$W and an excitation wavelength of 875~nm are used, and the PL image is extracted at an emission wavelength of 1518~nm. The scale bars are 1~$\mu$m. (c) A PL excitation map taken with an excitation power of 1~$\mu$W. (d) Polarization dependence of PL intensity. The line is a fit \cite{Moritsubo:2010}. (e) Excitation power dependence of PL intensity under CW (blue circles) and pulsed excitation (red squares). (f) An autocorrelation histogram taken at an excitation power of 15~nW.}
\end{figure}

PL measurements are performed with a home-built sample-scanning confocal microscopy system \cite{Ishii:2015}. We use a wavelength-tunable Ti:sapphire laser where the output can be switched between continuous-wave (CW) and $\sim$100-fs pulses with a repetition rate of 76~MHz. An excitation laser beam with power $P$ is focused onto the sample by an objective lens with a numerical aperture of 0.85 and a focal length of 1.8~mm, resulting in a $1/e^2$ diameter of $\sim$1~$\mu$m. PL and the reflected beam are collected by the same objective lens and separated by a dichroic filter. A Si photodiode detects the reflected beam for imaging, while a translating mirror is used to switch between PL spectroscopy and correlation measurements. PL spectra are measured with an InGaAs photodiode array attached to a spectrometer, and photon correlation measurements are performed using a Hanbury-Brown-Twiss (HBT) setup with a 50:50 beam splitter and fiber-coupled InGaAs single photon detectors. All measurements are performed at room temperature in dry air.

For characterizing the air-suspended nanotubes, we perform PL spectroscopy measurements under CW excitation \cite{Ishii:2015}. Line scans along the trenches are used to locate the suspended nanotubes, and we take reflectivity and PL images to confirm that it is fully suspended [Figs.~\ref{Fig1}(a) and \ref{Fig1}(b)]. The PL image shows a smooth spatial profile, indicating that the suspended nanotube is defect-free and does not contain any quenching sites or trapping sites \cite{Yoshikawa:2010, Harrah:2011a}. Next, we perform PL excitation spectroscopy to identify the chirality $(n,m)$ and to confirm that the nanotube is not bundled [Fig.~\ref{Fig1}(c)]. We find that the chirality of this nanotube is (10,9) with $E_{11}$ and $E_{22}$ resonances at 1518~nm and 877~nm, respectively. Polarization dependence of PL intensity is then measured to determine the angle of the nanotube [Fig.~\ref{Fig1}(d)], and the suspended length $L=2.58$~$\mu$m is obtained using the angle and the trench width \cite{Moritsubo:2010}. After setting the excitation resonant to $E_{22}$ and the laser polarization parallel to the tube axis, we take 50 PL spectra repeatedly to check that the nanotube does not show intermittency or spectral diffusion. Figure~\ref{Fig1}(e) shows the excitation power dependence, and it is confirmed that the PL intensity under CW excitation is proportional to $\sqrt[3]{P}$ at high powers \cite{Ishii:2015}. We have also measured the power dependence under pulsed excitation, where PL saturation results from more efficient EEA \cite{Murakami:2009prb2, Xiao:2010}.

Photon antibunching characteristics of the nanotubes are investigated by correlation measurements under pulsed excitation using the HBT setup. Because of the low efficiency of the single photon detectors, we need data accumulation times of up to 50 hours to obtain an autocorrelation histogram. Defocusing and misalignment of the excitation beam spot due to sample drift is avoided by tracking the nanotube position using three-axis sample scans performed every hour. Figure~\ref{Fig1}(f) is an autocorrelation histogram collected from the same nanotube presented in Figs.~\ref{Fig1}(a-e), clearly showing photon antibunching.

\begin{figure}
\includegraphics{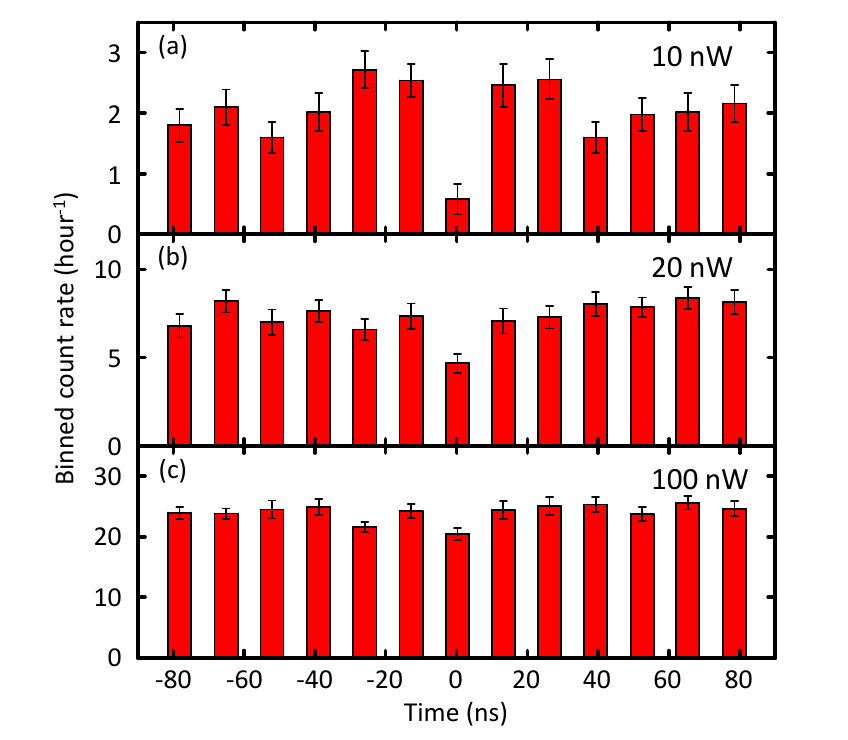}
\caption{
\label{Fig2} Binned count rate histograms taken with excitation powers of (a) 10~nW, (b) 20~nW, and (c) 100~nW. The binning width is 10 pixels which correspond to 2.0~ns. The error bars indicate standard error of the mean.}
\end{figure}

We evaluate photon antibunching from the autocorrelation histograms by subtracting the dark counts and binning each peak. Figure~\ref{Fig2} shows the binned count rate histograms of the (10,9) nanotube taken at various $P$, and we find that antibunching is clearer at lower $P$. We compute $g^{(2)}(0)$ by taking the ratio of the peak count at $\tau=0$ to the average of the other peak counts, and obtain $g^{(2)}(0)=0.27 \pm 0.12$, $0.62 \pm 0.07$, and $0.84 \pm 0.04$ at $P=10$, 20, and 100~nW, respectively.

\begin{figure}
\includegraphics{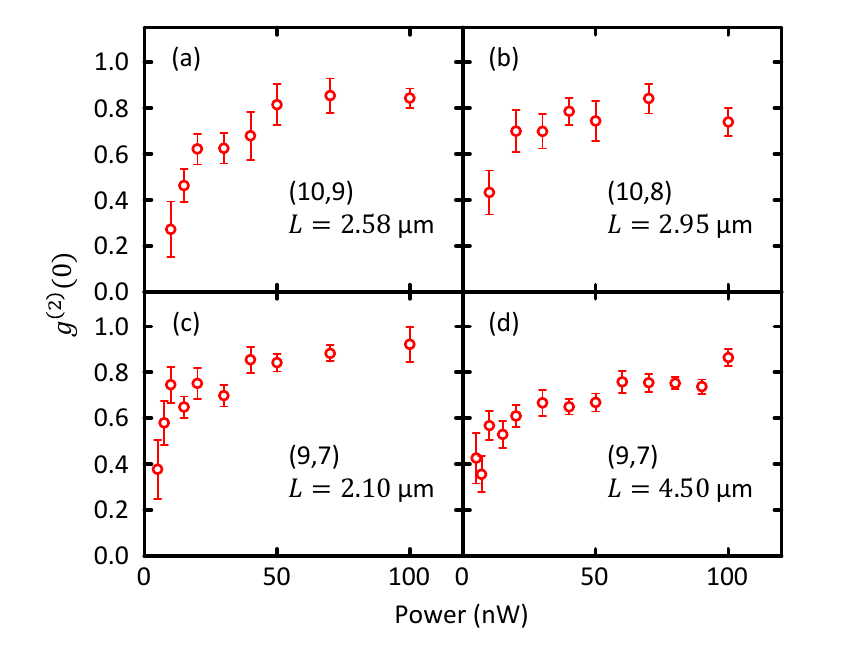}
\caption{
\label{Fig3} Excitation power dependence of $g^{(2)}(0)$ for four different nanotubes. (a) (10,9) nanotube with $L=2.58$~$\mu$m and emission wavelength of 1518~nm. (b) (10,8) nanotube with $L=2.95$~$\mu$m and emission wavelength of 1426~nm. (c) (9,7) nanotube with $L=2.10$~$\mu$m and emission wavelength of 1291~nm. (d) (9,7) nanotube with $L=4.50$~$\mu$m and emission wavelength of 1290~nm. The error bars show standard error of the mean.}
\end{figure}

Similar measurements are also performed on a (10,8) tube as well as two (9,7) tubes with different lengths, and the evaluated $g^{(2)}(0)$ is plotted as a function of $P$ in Fig.~\ref{Fig3}. We find that $g^{(2)}(0)$ decreases for lower excitation powers for all the nanotubes, although extrapolation down to $P=0$ does not seem to reach $g^{(2)}(0)=0$ except for the (10,9) nanotube. This minimum value
\begin{equation}
g_\mathrm{min} = \lim_{P \to 0} g^{(2)}(0)
\end{equation}
is less than 0.5 for all the tubes, indicating single photon emission. At high powers, $g^{(2)}(0)$ saturates at a value smaller than 1, which we define as
\begin{equation}
g_\mathrm{max} = \lim_{P \to \infty} g^{(2)}(0).
\end{equation}
In the data shown in Fig.~\ref{Fig3}, the values of $g_\mathrm{min}$ and $g_\mathrm{max}$ seem to be different among the nanotubes, suggesting that the single photon generation process is influenced by the chirality and the suspended length.

\section{Monte Carlo simulations of single photon generation by mobile excitons}
\begin{figure*}
\includegraphics{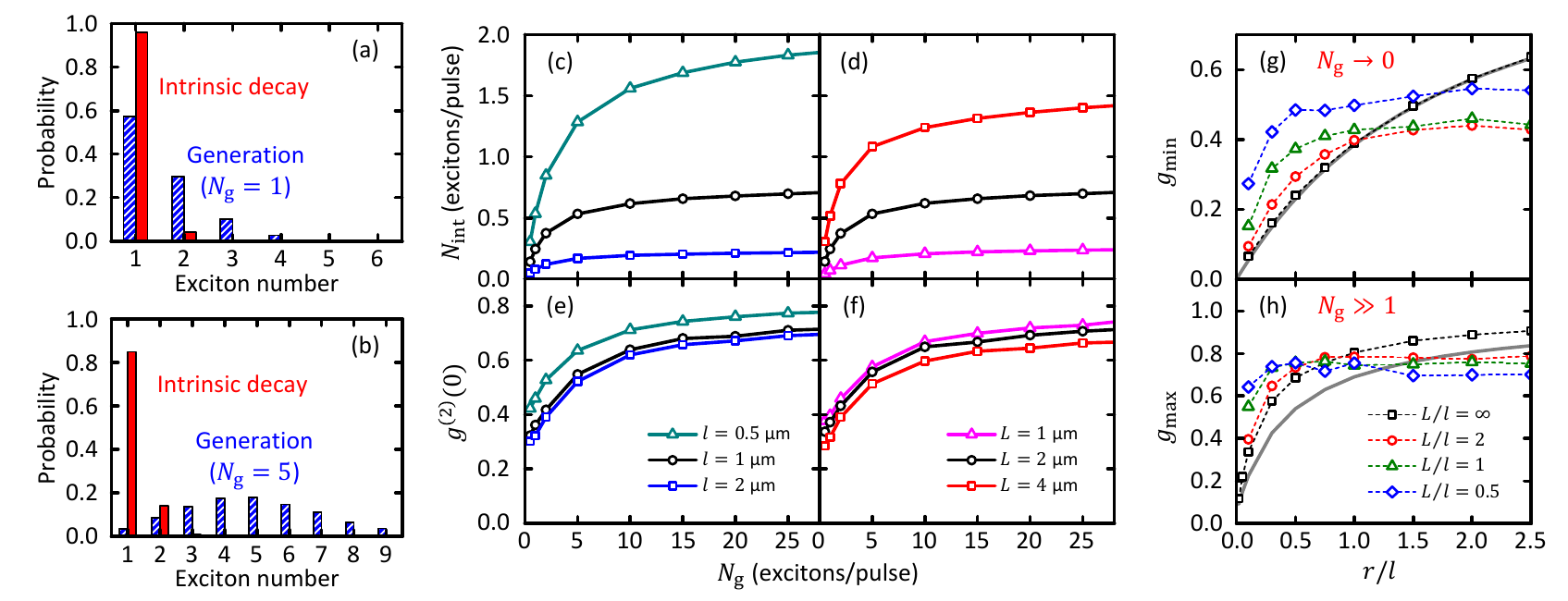}
\caption{
\label{Fig4} (a) and (b) Probability distributions obtained from simulations at $N_\mathrm{g}=1$ and $N_\mathrm{g}=5$, respectively. Blue striped bars indicate $n_\mathrm{g}$, and red filled bars represent $n_\mathrm{int}$. (c) and (e) $N_\mathrm{g}$ dependence of $N_\mathrm{int}$ and $g^{(2)}(0)$, respectively, with $r=0.5$~$\mu$m and $L=2$~$\mu$m. $l=0.5$~$\mu$m (green triangles), $l=1$~$\mu$m (black circles), and $l=2$~$\mu$m (blue squares) are used. (d) and (f) $N_\mathrm{g}$ dependence of $N_\mathrm{int}$ and $g^{(2)}(0)$, respectively, with $l=1$~$\mu$m and $r=0.5$~$\mu$m. $L=1$~$\mu$m (magenta triangles), $L=2$~$\mu$m (black circles), and $L=4$~$\mu$m (red squares) are used. (g) and (h) $g_\mathrm{min}$ and $g_\mathrm{max}$, respectively, obtained from simulations and calculation. Simulations are performed with infinitely long $L$ (black squares), $L/l=2$ (red circles), $L/l=1$ (green triangles), and $L/l=0.5$ (blue diamonds). Gray solid lines are obtained by Eq.~(\ref{gmin}) for (g) and Eq.~(\ref{gmax}) for (h).}
\end{figure*}

In order to understand the power dependence as well as the differences between the four tubes, we conduct Monte Carlo simulations of diffusion-driven single photon generation. We use the same method as in Ref.~\cite{Ishii:2015}, and modify the excitation process. The excitons generated by the laser pulse obey the Poisson distribution
\begin{equation}
p_\mathrm{g} (n_\mathrm{g} ) =\frac{{N_\mathrm{g}}^{n_\mathrm{g} }}{ n_\mathrm{g} !} \mathrm{exp}⁡(-N_\mathrm{g}),
\end{equation}
where $n_\mathrm{g}$ is the number of excitons and $N_\mathrm{g} = \left< n_\mathrm{g} \right>$ is the average. After letting all excitons decay, we record the number of excitons that went through the intrinsic decay process as $n_\mathrm{int}$, and by repeating the simulation trials,
\begin{equation}
g^{(2)}(0) = \frac{V_\mathrm{int} - N_\mathrm{int}}{{N_\mathrm{int}}^2} +1
\end{equation}
is obtained where $N_\mathrm{int} = \left< n_\mathrm{int} \right>$ and $V_\mathrm{int} = \left< {n_\mathrm{int}}^2 \right> - { \left< n_\mathrm{int} \right>}^2$ are the average and variance of $n_\mathrm{int}$, respectively. The excitation power in the experiments would be proportional to $N_\mathrm{g}$, while PL intensity corresponds to $N_\mathrm{int}$. We note that $g^{(2)}(0)$ evaluated from simulations can be directly compared to the experimental values, although the detected photon counts are affected by radiative quantum efficiency of excitons, photon collection efficiency in the optical system, and detection efficiency of the single photon detectors \cite{Nair:2011}.

We have performed the simulations for low ($N_\mathrm{g}=1$) and high ($N_\mathrm{g}=5$) generation rates, and the probability distributions of exciton numbers are shown in Figs.~\ref{Fig4}(a) and \ref{Fig4}(b), respectively. Here we use exciton diffusion length $l=1$~$\mu$m and excitation laser $1/e^2$ radius $r=0.5$~$\mu$m as well as $L=2$~$\mu$m. $n_\mathrm{g}$ follows the Poisson distribution as implemented, while $n_\mathrm{int}$ takes smaller values because of efficient EEA. In the case of a low generation rate [Fig.~\ref{Fig4}(a)], only a single intrinsic decay event occurs for most of the trials, and $g^{(2)}(0)$ is as small as 0.33. For a higher generation rate [Fig.~\ref{Fig4}(b)], the fraction of $n_\mathrm{int} \ge 2$ increases and thus $g^{(2)}(0)$ becomes larger. It is worth mentioning that antibunching and the sub-Poissonian distribution of $n_\mathrm{int}$ are a direct consequence of EEA, and $g^{(2)}(0)=1$ would result if EEA is absent. These simulation results indicate that EEA of mobile excitons in micron-long tubes is efficient enough to cause antibunching.

The $N_\mathrm{g}$ dependence is investigated in more detail by performing the simulations with various $l$ and $L$. In Figs.~\ref{Fig4}(c) and \ref{Fig4}(d), $N_\mathrm{int}$ is plotted as a function of $N_\mathrm{g}$, showing saturation at high $N_\mathrm{g}$ as we experimentally observed in the excitation power dependence of PL intensity under pulsed excitation [Fig.~\ref{Fig1}(e)]. The $N_\mathrm{g}$ dependence of $g^{(2)}(0)$ obtained from the same simulations are plotted in Figs.~\ref{Fig4}(e) and \ref{Fig4}(f), where $g^{(2)}(0)$ is smallest at the lowest $N_\mathrm{g}$ and saturates at high $N_\mathrm{g}$. This behavior is also in agreement with our experimental results (Fig.~\ref{Fig3}), showing that the simulations can reproduce the experimental power dependence of PL intensity and $g^{(2)}(0)$ simultaneously.

Next, we consider the influence of $l$ and $L$. With longer $l$, $N_\mathrm{int}$ becomes lower because the decay rates of both EEA and end quenching increases [Fig.~\ref{Fig4}(c)]. $g^{(2)}(0)$ also decreases slightly for longer $l$ [Fig.~\ref{Fig4}(e)], suggesting that the contribution of EEA is larger than end quenching. Among the four nanotubes shown in Fig.~\ref{Fig3}, (10,9) nanotubes have $l>3~\mu$m, much longer compared to $l=1.13$~$\mu$m for (9,7) and $l=0.47$~$\mu$m for (10,8) nanotubes \cite{Ishii:2015}, and indeed the (10,9) nanotube exhibits the lowest $g_\mathrm{min}$. In comparison, the effects of $L$ can be more beneficial for single photon generation. In the case of longer $L$, higher $N_\mathrm{int}$ and lower $g^{(2)}(0)$ are obtained at the same time [Figs.~\ref{Fig4}(d) and \ref{Fig4}(f)]. This is because the influence of end quenching decreases as $L$ increases, and then the fraction of excitons that are involved in EEA increases. By comparing the two (9,7) nanotubes with different $L$ [Figs.~\ref{Fig3}(c) and \ref{Fig3}(d)], the longer one yields higher PL intensity and lower $g^{(2)}(0)$ at all excitation powers, which agrees with our explanation. These observations suggest that longer $l$ and $L$ would lead to even lower $g_\mathrm{min}$.

\section{First-passage approach}
In order to find ways to reduce $g_\mathrm{min}$, we look for an analytical expression of $g_\mathrm{min}$ in ideal situations where end quenching is negligible. $g_\mathrm{min}$ is given by $\eta_\mathrm{2ex} / \eta_\mathrm{1ex}$ \cite{Nair:2011}, where $\eta_\mathrm{2ex}$ and $\eta_\mathrm{1ex}$ are emission efficiencies for the cases when the number of excitons decreases from 2 to 1, and from 1 to 0, respectively. As emission from the two-exciton state can only occur when one of the excitons go through radiative decay without any EEA, $\eta_\mathrm{2ex}$ can be written as $\eta_\mathrm{1ex} \overline{S_\mathrm{EEA}}$ using the average EEA survival probability $\overline{S_\mathrm{EEA}}$, and therefore $g_\mathrm{min} = \overline{S_\mathrm{EEA}}$.

An expression for $\overline{S_\mathrm{EEA}}$ is obtained by using first passage approach. We start by deriving the EEA survival probability $S_\mathrm{EEA}(x_0)$ for a pair of excitons with an initial separation distance $x_0$. As relative diffusion of two excitons is equivalent to the case where one is fixed and the other has twice the diffusivity \cite{Redner}, $S_\mathrm{EEA}(x_0)$ is equal to the probability for intrinsic decay to occur without ever arriving at the fixed point separated by $x_0$. First, we consider the exciton density distribution without the EEA process. A steady-state one-dimensional diffusion equation with exciton generation at $x=0$ is given by \cite{Moritsubo:2010}
\begin{equation}
2D \frac{d^2 n(x)}{dx^2} - \frac{2}{\tau} n(x) + G_0 \delta (x) = 0,
\label{DiffEq1}
\end{equation}
where $n(x)$ is the exciton density, $x$ is the position on the nanotube, $D$ is the diffusivity, $\tau$ is the intrinsic lifetime, $G_0$ is the exciton generation rate, and $\delta(x)$ is the Dirac delta function. In Eq.~(\ref{DiffEq1}), diffusivity of $2D$ is used as mentioned above, and a lifetime of $\tau/2$ is used because either one of the two excitons can recombine. Using the relation $l=\sqrt{D\tau}$, Eq.~(\ref{DiffEq1}) can be written as
\begin{equation}
l^2 \frac{d^2 n(x)}{dx^2} - n(x) + N_0 \delta (x) = 0,
\label{DiffEq2}
\end{equation}
where $N_0=G_0\tau/2$ is a constant. By solving Eq.~(\ref{DiffEq2}) with the boundary conditions $n(\pm \infty)=0$,
\begin{equation}
n(x) = \frac{N_0}{2l} \mathrm{exp}\left(-\frac{|x|}{l}\right)
\label{n_x}
\end{equation}
is obtained, where the total number of excitons in this distribution is $\int_{-\infty}^{\infty} n(x) dx = N_0$. Next, we calculate the amount of excitons that have diffused through the fixed point $x=x_0>0$ at least once. The number of excitons in the region beyond the fixed point is $\int_{x_0}^{\infty} n(x) dx$, and these excitons have obviously traveled farther than $x=x_0$. Based on the reflection principle \cite{Redner}, the same amount of excitons should have diffused back to $x<x_0$ after reaching the fixed point. The total number of excitons that have ever visited the fixed point is therefore given by
\begin{equation}
N_\mathrm{1} = 2\int_{x_0}^{\infty} n(x) dx = N_0 \mathrm{exp}\left(-\frac{x_0}{l}\right).
\end{equation}
Finally, by calculating the fraction of excitons that have not passed the fixed point,
\begin{equation}
S_\mathrm{EEA}(x_0) = \frac{N_0 - N_\mathrm{1}}{N_0} = 1-\mathrm{exp}\left(-\frac{x_0}{l}\right)
\end{equation}
is obtained.

We use this probability to compute $\overline{S_\mathrm{EEA}}$. The initial positions of the excitons are stochastically determined by the Gaussian profile of the excitation laser, and the probability distribution of $x_0$ is given by $p_\mathrm{d}(x_0) = \frac{2}{\sqrt{\pi r^2}}\mathrm{exp}⁡(-{x_0}^2/r^2)$, where the variance of $x_0$ is doubled from that of initial positions of the excitons. Finally,
\begin{equation}
\begin{split}
\overline{S_\mathrm{EEA}}
&= \int_{0}^{\infty} S_\mathrm{EEA}(x_0) p_\mathrm{d}(x_0) dx_0\\
&= 1 - \mathrm{exp}\left[ \left( \frac{r}{2l} \right)^2 \right]\mathrm{erfc}\left(\frac{r}{2l}\right),
\label{gmin}
\end{split}
\end{equation}
where $\mathrm{erfc}(x) = 1- \frac{2}{\sqrt{\pi}} \int_0^x \exp (-y^2) dy$ is the complimentary error function.

In Fig.~\ref{Fig4}(g), $g_\mathrm{min}=\overline{S_\mathrm{EEA}}$ calculated by Eq.~(\ref{gmin}) is shown as a gray solid line, indicating that $g_\mathrm{min}$ can be lowered by reducing $r/l$ and that it can reach a value near zero. We also obtain $g_\mathrm{min}$ from simulations by considering a situation where $N_\mathrm{g} \to 0$, and it is confirmed that Eq.~(\ref{gmin}) matches the simulation results with an infinitely long $L$ [black squares in Fig.~\ref{Fig4}(g)]. This implies that high-purity single photon emission can in principle be obtained from mobile excitons in long carbon nanotubes under appropriate conditions. Although longer $l$ is not easily obtained in experiments, $r$ can be made a few orders of magnitude smaller by utilizing near-field techniques \cite{Rauhut:2012, Liao:2016}, and therefore $g_\mathrm{min} \sim 0$ should be feasible. This result is different from that suggested by a model in Ref.~\cite{Ma:2015PRL}, where the degree of antibunching remains constant at $g_\mathrm{min}\gtrsim0.4$ even for small $r/l$.

In cases where $L$ is finite, simulation results deviate from Eq.~(\ref{gmin}) due to the influence of end quenching. In the region of $r<L$, efficiency of end quenching increases as $r$ increases, reducing EEA probability and increasing $g_\mathrm{min}$. Such an influence of end quenching is larger for shorter $L$ and longer $l$, and therefore $g_\mathrm{min}$ becomes large for smaller $L/l$. It is interesting to note that longer nanotubes are advantageous for producing single photons, whereas conventional single photon emitters are nanometer-scale or less. We may compare the simulation results in this region to experimental data as $r/l$ are within 0.2 to 1.1 while $L/l$ range from 0.9 to 6.3, assuming $l=3$~$\mu$m for the (10,9) nanotube. The experimentally obtained minimum values of $g^{(2)}(0)$ for all four tubes are consistent with the simulation results, falling within a reasonable range of 0.27 to 0.43. Now considering the region of $r>L$, $L$ limits the initial distance between two excitons, resulting in the plateau since the situation does not change for larger $r$. We believe that the plateau of $g_\mathrm{min}$ at $\sim0.4$ correspond to the limiting value obtained under uniform excitation within the diffusion length \cite{Ma:2015PRL}.

Finally, we look into $g_\mathrm{max}$ by performing simulations with $N_\mathrm{g}=1000$, where $g^{(2)}(0)$ and $N_\mathrm{int}$ are completely saturated. When $L/l$ is large enough for end quenching to be negligible, $g_\mathrm{max}$ decreases as $r/l$ decreases [black squares in Fig.~\ref{Fig4}(h)]. This behavior can be qualitatively explained by noting that
\begin{equation}
g_\mathrm{max}=\frac{m-1}{m}
\label{gmax}
\end{equation}
for $m$ independent emitters \cite{Tinnefeld:2001}, where we use the saturated $N_\mathrm{int}$ as the number of emitters in the nanotube [gray line in Fig.~\ref{Fig4}(h)]. This model suggests that $g_\mathrm{max}$ can also be lowered by reducing $r/l$, which would allow for higher single photon emission rates. For finite $L$, effects of end quenching come into play as in the case with $g_\mathrm{min}$. Within the region of $r<L$, end quenching becomes more efficient as $r$ increases, resulting in larger $g_\mathrm{max}$. $L$ also affects the efficiency of end quenching within this region, and $g_\mathrm{max}$ increases for smaller $L/l$. As $r$ gets longer beyond $L$, the number of independent emitters is limited by $L$, and therefore $g_\mathrm{max}$ does not change for larger $r$. This picture can explain why smaller $L/l$ yields lower $g_\mathrm{max}$, despite the increase in end quenching.

\section{Conclusion}
In conclusion, we demonstrate room-temperature single photon emission from micron-long air-suspended carbon nanotubes, and we show that high-purity single photons can be generated by mobile excitons in carbon nanotubes. Photon antibunching is observed in correlation measurements on individual nanotubes with suspended lengths over 2~$\mu$m, where we find that $g^{(2)}(0)$ decreases as excitation power is lowered. We have measured photon statistics on four nanotubes with different chiralities and suspended lengths, while Monte Carlo simulations reveal that the characteristics of $g^{(2)}(0)$ can be explained by the competition between EEA and end quenching processes. We derive an analytical expression for the minimum value of $g^{(2)}(0)$ attained at the low excitation power limit, which indicates that the purity of single photon emission can in principle be improved toward $g^{(2)}(0)\sim0$ by reducing the excitation spot size within a long nanotube. Such a micron-scale platform for quantum light sources allows integration into electronic devices \cite{Higashide:2017} or photonic structures \cite{Miura:2014}, opening up a pathway to devices with additional functionality and flexibility.

\begin{acknowledgments}
Work supported by JSPS (KAKENHI JP16H05962) and MEXT (Photon Frontier Network Program, Nanotechnology Platform). T.U. is supported by ALPS and JSPS Research Fellowship.
\end{acknowledgments}


\begin{thebibliography}{41}%
\makeatletter
\providecommand \@ifxundefined [1]{%
 \@ifx{#1\undefined}
}%
\providecommand \@ifnum [1]{%
 \ifnum #1\expandafter \@firstoftwo
 \else \expandafter \@secondoftwo
 \fi
}%
\providecommand \@ifx [1]{%
 \ifx #1\expandafter \@firstoftwo
 \else \expandafter \@secondoftwo
 \fi
}%
\providecommand \natexlab [1]{#1}%
\providecommand \enquote  [1]{``#1''}%
\providecommand \bibnamefont  [1]{#1}%
\providecommand \bibfnamefont [1]{#1}%
\providecommand \citenamefont [1]{#1}%
\providecommand \href@noop [0]{\@secondoftwo}%
\providecommand \href [0]{\begingroup \@sanitize@url \@href}%
\providecommand \@href[1]{\@@startlink{#1}\@@href}%
\providecommand \@@href[1]{\endgroup#1\@@endlink}%
\providecommand \@sanitize@url [0]{\catcode `\\12\catcode `\$12\catcode
  `\&12\catcode `\#12\catcode `\^12\catcode `\_12\catcode `\%12\relax}%
\providecommand \@@startlink[1]{}%
\providecommand \@@endlink[0]{}%
\providecommand \url  [0]{\begingroup\@sanitize@url \@url }%
\providecommand \@url [1]{\endgroup\@href {#1}{\urlprefix }}%
\providecommand \urlprefix  [0]{URL }%
\providecommand \Eprint [0]{\href }%
\providecommand \doibase [0]{https://doi.org/}%
\providecommand \selectlanguage [0]{\@gobble}%
\providecommand \bibinfo  [0]{\@secondoftwo}%
\providecommand \bibfield  [0]{\@secondoftwo}%
\providecommand \translation [1]{[#1]}%
\providecommand \BibitemOpen [0]{}%
\providecommand \bibitemStop [0]{}%
\providecommand \bibitemNoStop [0]{.\EOS\space}%
\providecommand \EOS [0]{\spacefactor3000\relax}%
\providecommand \BibitemShut  [1]{\csname bibitem#1\endcsname}%
\let\auto@bib@innerbib\@empty
\bibitem [{\citenamefont {O'Brien}\ \emph {et~al.}(2009)\citenamefont
  {O'Brien}, \citenamefont {Furusawa},\ and\ \citenamefont
  {Vu\v{c}kovic}}]{OBrien:2009}%
  \BibitemOpen
  \bibfield  {author} {\bibinfo {author} {\bibfnamefont {J.~L.}\ \bibnamefont
  {O'Brien}}, \bibinfo {author} {\bibfnamefont {A.}~\bibnamefont {Furusawa}}, \
  and\ \bibinfo {author} {\bibfnamefont {J.}~\bibnamefont {Vu\v{c}kovic}},\
  }\bibfield  {title} {\bibinfo {title} {Photonic quantum technologies},\
  }\href {\doibase 10.1038/nphoton.2009.229} {\bibfield  {journal} {\bibinfo
  {journal} {Nat. Photon.}\ }\textbf {\bibinfo {volume} {3}},\ \bibinfo {pages}
  {687} (\bibinfo {year} {2009})}\BibitemShut {NoStop}%
\bibitem [{\citenamefont {Kimble}\ \emph {et~al.}(1977)\citenamefont {Kimble},
  \citenamefont {Dagenais},\ and\ \citenamefont {Mandel}}]{kimble:1977}%
  \BibitemOpen
  \bibfield  {author} {\bibinfo {author} {\bibfnamefont {H.~J.}\ \bibnamefont
  {Kimble}}, \bibinfo {author} {\bibfnamefont {M.}~\bibnamefont {Dagenais}}, \
  and\ \bibinfo {author} {\bibfnamefont {L.}~\bibnamefont {Mandel}},\
  }\bibfield  {title} {\bibinfo {title} {Photon antibunching in resonance
  fluorescence},\ }\href {\doibase 10.1103/PhysRevLett.39.691} {\bibfield
  {journal} {\bibinfo  {journal} {Phys. Rev. Lett.}\ }\textbf {\bibinfo
  {volume} {39}},\ \bibinfo {pages} {691} (\bibinfo {year} {1977})}\BibitemShut
  {NoStop}%
\bibitem [{\citenamefont {Diedrich}\ and\ \citenamefont
  {Walther}(1987)}]{diedrich:1987}%
  \BibitemOpen
  \bibfield  {author} {\bibinfo {author} {\bibfnamefont {F.}~\bibnamefont
  {Diedrich}}\ and\ \bibinfo {author} {\bibfnamefont {H.}~\bibnamefont
  {Walther}},\ }\bibfield  {title} {\bibinfo {title} {Nonclassical radiation of
  a single stored ion},\ }\href {\doibase 10.1103/PhysRevLett.58.203}
  {\bibfield  {journal} {\bibinfo  {journal} {Phys. Rev. Lett.}\ }\textbf
  {\bibinfo {volume} {58}},\ \bibinfo {pages} {203} (\bibinfo {year}
  {1987})}\BibitemShut {NoStop}%
\bibitem [{\citenamefont {Basch\'e}\ \emph {et~al.}(1992)\citenamefont
  {Basch\'e}, \citenamefont {Moerner}, \citenamefont {Orrit},\ and\
  \citenamefont {Talon}}]{basche:1992}%
  \BibitemOpen
  \bibfield  {author} {\bibinfo {author} {\bibfnamefont {T.}~\bibnamefont
  {Basch\'e}}, \bibinfo {author} {\bibfnamefont {W.~E.}\ \bibnamefont
  {Moerner}}, \bibinfo {author} {\bibfnamefont {M.}~\bibnamefont {Orrit}}, \
  and\ \bibinfo {author} {\bibfnamefont {H.}~\bibnamefont {Talon}},\ }\bibfield
   {title} {\bibinfo {title} {Photon antibunching in the fluorescence of a
  single dye molecule trapped in a solid},\ }\href {\doibase 10.1103/PhysRevLett.69.1516} {\bibfield  {journal} {\bibinfo  {journal}
  {Phys. Rev. Lett.}\ }\textbf {\bibinfo {volume} {69}},\ \bibinfo {pages}
  {1516} (\bibinfo {year} {1992})}\BibitemShut {NoStop}%
\bibitem [{\citenamefont {Lounis}\ and\ \citenamefont
  {Moerner}(2000)}]{lounis:2000}%
  \BibitemOpen
  \bibfield  {author} {\bibinfo {author} {\bibfnamefont {B.}~\bibnamefont
  {Lounis}}\ and\ \bibinfo {author} {\bibfnamefont {W.~E.}\ \bibnamefont
  {Moerner}},\ }\bibfield  {title} {\bibinfo {title} {Single photons on demand
  from a single molecule at room temperature},\ }\href {\doibase 10.1038/35035032} {\bibfield  {journal} {\bibinfo  {journal} {Nature}\
  }\textbf {\bibinfo {volume} {407}},\ \bibinfo {pages} {491} (\bibinfo {year}
  {2000})}\BibitemShut {NoStop}%
\bibitem [{\citenamefont {Michler}\ \emph {et~al.}(2000)\citenamefont
  {Michler}, \citenamefont {Imamo\u{g}lu}, \citenamefont {Mason}, \citenamefont
  {Carson}, \citenamefont {Strouse},\ and\ \citenamefont
  {Buratto}}]{michler:2000nature}%
  \BibitemOpen
  \bibfield  {author} {\bibinfo {author} {\bibfnamefont {P.}~\bibnamefont
  {Michler}}, \bibinfo {author} {\bibfnamefont {A.}~\bibnamefont
  {Imamo\u{g}lu}}, \bibinfo {author} {\bibfnamefont {M.~D.}\ \bibnamefont
  {Mason}}, \bibinfo {author} {\bibfnamefont {P.~J.}\ \bibnamefont {Carson}},
  \bibinfo {author} {\bibfnamefont {G.~F.}\ \bibnamefont {Strouse}}, \ and\
  \bibinfo {author} {\bibfnamefont {S.~K.}\ \bibnamefont {Buratto}},\
  }\bibfield  {title} {\bibinfo {title} {Quantum correlation among photons from
  a single quantum dot at room temperature},\ }\href {\doibase 10.1038/35023100} {\bibfield  {journal} {\bibinfo  {journal} {Nature}\
  }\textbf {\bibinfo {volume} {406}},\ \bibinfo {pages} {968} (\bibinfo {year}
  {2000})}\BibitemShut {NoStop}%
\bibitem [{\citenamefont {Choi}\ \emph {et~al.}(2014)\citenamefont {Choi},
  \citenamefont {Johnson}, \citenamefont {Castelletto}, \citenamefont
  {Ton-That}, \citenamefont {Phillips},\ and\ \citenamefont
  {Aharonovich}}]{choi:2014}%
  \BibitemOpen
  \bibfield  {author} {\bibinfo {author} {\bibfnamefont {S.}~\bibnamefont
  {Choi}}, \bibinfo {author} {\bibfnamefont {B.~C.}\ \bibnamefont {Johnson}},
  \bibinfo {author} {\bibfnamefont {S.}~\bibnamefont {Castelletto}}, \bibinfo
  {author} {\bibfnamefont {C.}~\bibnamefont {Ton-That}}, \bibinfo {author}
  {\bibfnamefont {M.~R.}\ \bibnamefont {Phillips}}, \ and\ \bibinfo {author}
  {\bibfnamefont {I.}~\bibnamefont {Aharonovich}},\ }\bibfield  {title}
  {\bibinfo {title} {Single photon emission from {ZnO} nanoparticles},\ }\href
  {\doibase 10.1063/1.4872268} {\bibfield  {journal} {\bibinfo  {journal}
  {Appl. Phys. Lett.}\ }\textbf {\bibinfo {volume} {104}},\ \bibinfo {pages}
  {261101} (\bibinfo {year} {2014})}\BibitemShut {NoStop}%
\bibitem [{\citenamefont {Kurtsiefer}\ \emph {et~al.}(2000)\citenamefont
  {Kurtsiefer}, \citenamefont {Mayer}, \citenamefont {Zarda},\ and\
  \citenamefont {Weinfurter}}]{Kurtsiefer:2000}%
  \BibitemOpen
  \bibfield  {author} {\bibinfo {author} {\bibfnamefont {C.}~\bibnamefont
  {Kurtsiefer}}, \bibinfo {author} {\bibfnamefont {S.}~\bibnamefont {Mayer}},
  \bibinfo {author} {\bibfnamefont {P.}~\bibnamefont {Zarda}}, \ and\ \bibinfo
  {author} {\bibfnamefont {H.}~\bibnamefont {Weinfurter}},\ }\bibfield  {title}
  {\bibinfo {title} {Stable solid-state source of single photons},\ }\href
  {\doibase 10.1103/PhysRevLett.85.290} {\bibfield  {journal} {\bibinfo
  {journal} {Phys. Rev. Lett.}\ }\textbf {\bibinfo {volume} {85}},\ \bibinfo
  {pages} {290} (\bibinfo {year} {2000})}\BibitemShut {NoStop}%
\bibitem [{\citenamefont {Castelletto}\ \emph {et~al.}(2014)\citenamefont
  {Castelletto}, \citenamefont {Johnson}, \citenamefont {Iv{\'a}dy},
  \citenamefont {Stavrias}, \citenamefont {Umeda}, \citenamefont {Gali},\ and\
  \citenamefont {Ohshima}}]{castelletto:2014}%
  \BibitemOpen
  \bibfield  {author} {\bibinfo {author} {\bibfnamefont {S.}~\bibnamefont
  {Castelletto}}, \bibinfo {author} {\bibfnamefont {B.~C.}\ \bibnamefont
  {Johnson}}, \bibinfo {author} {\bibfnamefont {V.}~\bibnamefont {Iv{\'a}dy}},
  \bibinfo {author} {\bibfnamefont {N.}~\bibnamefont {Stavrias}}, \bibinfo
  {author} {\bibfnamefont {T.}~\bibnamefont {Umeda}}, \bibinfo {author}
  {\bibfnamefont {A.}~\bibnamefont {Gali}}, \ and\ \bibinfo {author}
  {\bibfnamefont {T.}~\bibnamefont {Ohshima}},\ }\bibfield  {title} {\bibinfo
  {title} {A silicon carbide room-temperature single-photon source},\ }\href
  {\doibase 10.1038/nmat3806} {\bibfield  {journal} {\bibinfo  {journal} {Nat.
  Mater.}\ }\textbf {\bibinfo {volume} {13}},\ \bibinfo {pages} {151} (\bibinfo
  {year} {2014})}\BibitemShut {NoStop}%
\bibitem [{\citenamefont {Benyoucef}\ \emph {et~al.}(2013)\citenamefont
  {Benyoucef}, \citenamefont {Yacob}, \citenamefont {Reithmaier}, \citenamefont
  {Kettler},\ and\ \citenamefont {Michler}}]{Benyoucef:2013}%
  \BibitemOpen
  \bibfield  {author} {\bibinfo {author} {\bibfnamefont {M.}~\bibnamefont
  {Benyoucef}}, \bibinfo {author} {\bibfnamefont {M.}~\bibnamefont {Yacob}},
  \bibinfo {author} {\bibfnamefont {J.~P.}\ \bibnamefont {Reithmaier}},
  \bibinfo {author} {\bibfnamefont {J.}~\bibnamefont {Kettler}}, \ and\
  \bibinfo {author} {\bibfnamefont {P.}~\bibnamefont {Michler}},\ }\bibfield
  {title} {\bibinfo {title} {Telecom-wavelength (1.5~$\mu$m) single-photon
  emission from {InP}-based quantum dots},\ }\href {\doibase 10.1063/1.4825106}
  {\bibfield  {journal} {\bibinfo  {journal} {Appl. Phys. Lett.}\ }\textbf
  {\bibinfo {volume} {103}},\ \bibinfo {pages} {162101} (\bibinfo {year}
  {2013})}\BibitemShut {NoStop}%
\bibitem [{\citenamefont {Takemoto}\ \emph {et~al.}(2010)\citenamefont
  {Takemoto}, \citenamefont {Nambu}, \citenamefont {Miyazawa}, \citenamefont
  {Wakui}, \citenamefont {Hirose}, \citenamefont {Usuki}, \citenamefont
  {Takatsu}, \citenamefont {Yokoyama}, \citenamefont {Yoshino}, \citenamefont
  {Tomita}, \citenamefont {Yorozu}, \citenamefont {Sakuma},\ and\ \citenamefont
  {Arakawa}}]{Takemoto:2010}%
  \BibitemOpen
  \bibfield  {author} {\bibinfo {author} {\bibfnamefont {K.}~\bibnamefont
  {Takemoto}}, \bibinfo {author} {\bibfnamefont {Y.}~\bibnamefont {Nambu}},
  \bibinfo {author} {\bibfnamefont {T.}~\bibnamefont {Miyazawa}}, \bibinfo
  {author} {\bibfnamefont {K.}~\bibnamefont {Wakui}}, \bibinfo {author}
  {\bibfnamefont {S.}~\bibnamefont {Hirose}}, \bibinfo {author} {\bibfnamefont
  {T.}~\bibnamefont {Usuki}}, \bibinfo {author} {\bibfnamefont
  {M.}~\bibnamefont {Takatsu}}, \bibinfo {author} {\bibfnamefont
  {N.}~\bibnamefont {Yokoyama}}, \bibinfo {author} {\bibfnamefont
  {K.}~\bibnamefont {Yoshino}}, \bibinfo {author} {\bibfnamefont
  {A.}~\bibnamefont {Tomita}}, \bibinfo {author} {\bibfnamefont
  {S.}~\bibnamefont {Yorozu}}, \bibinfo {author} {\bibfnamefont
  {Y.}~\bibnamefont {Sakuma}}, \ and\ \bibinfo {author} {\bibfnamefont
  {Y.}~\bibnamefont {Arakawa}},\ }\bibfield  {title} {\bibinfo {title}
  {Transmission experiment of quantum keys over 50 km using high-performance
  quantum-dot single-photon source at 1.5~$\mu$m wavelength},\ }\href {\doibase 10.1143/APEX.3.092802} {\bibfield  {journal} {\bibinfo  {journal} {Appl.
  Phys. Express}\ }\textbf {\bibinfo {volume} {3}},\ \bibinfo {pages} {092802}
  (\bibinfo {year} {2010})}\BibitemShut {NoStop}%
\bibitem [{\citenamefont {He}\ \emph {et~al.}(2015)\citenamefont {He},
  \citenamefont {Clark}, \citenamefont {Schaibley}, \citenamefont {He},
  \citenamefont {Chen}, \citenamefont {Wei}, \citenamefont {Ding},
  \citenamefont {Zhang}, \citenamefont {Yao}, \citenamefont {Xu}, \citenamefont
  {Lu},\ and\ \citenamefont {Pan}}]{He:2015}%
  \BibitemOpen
  \bibfield  {author} {\bibinfo {author} {\bibfnamefont {Y.-M.}\ \bibnamefont
  {He}}, \bibinfo {author} {\bibfnamefont {G.}~\bibnamefont {Clark}}, \bibinfo
  {author} {\bibfnamefont {J.~R.}\ \bibnamefont {Schaibley}}, \bibinfo {author}
  {\bibfnamefont {Y.}~\bibnamefont {He}}, \bibinfo {author} {\bibfnamefont
  {M.-C.}\ \bibnamefont {Chen}}, \bibinfo {author} {\bibfnamefont {Y.-J.}\
  \bibnamefont {Wei}}, \bibinfo {author} {\bibfnamefont {X.}~\bibnamefont
  {Ding}}, \bibinfo {author} {\bibfnamefont {Q.}~\bibnamefont {Zhang}},
  \bibinfo {author} {\bibfnamefont {W.}~\bibnamefont {Yao}}, \bibinfo {author}
  {\bibfnamefont {X.}~\bibnamefont {Xu}}, \bibinfo {author} {\bibfnamefont
  {C.-Y.}\ \bibnamefont {Lu}}, \ and\ \bibinfo {author} {\bibfnamefont {J.-W.}\
  \bibnamefont {Pan}},\ }\bibfield  {title} {\bibinfo {title} {Single quantum
  emitters in monolayer semiconductors},\ }\href {\doibase 10.1038/nnano.2015.75} {\bibfield  {journal} {\bibinfo  {journal} {Nat.
  Nanotech.}\ }\textbf {\bibinfo {volume} {10}},\ \bibinfo {pages} {497}
  (\bibinfo {year} {2015})}\BibitemShut {NoStop}%
\bibitem [{\citenamefont {Tran}\ \emph {et~al.}(2016)\citenamefont {Tran},
  \citenamefont {Bray}, \citenamefont {Ford}, \citenamefont {Toth},\ and\
  \citenamefont {Aharonovich}}]{Tran:2016}%
  \BibitemOpen
  \bibfield  {author} {\bibinfo {author} {\bibfnamefont {T.~T.}\ \bibnamefont
  {Tran}}, \bibinfo {author} {\bibfnamefont {K.}~\bibnamefont {Bray}}, \bibinfo
  {author} {\bibfnamefont {M.~J.}\ \bibnamefont {Ford}}, \bibinfo {author}
  {\bibfnamefont {M.}~\bibnamefont {Toth}}, \ and\ \bibinfo {author}
  {\bibfnamefont {I.}~\bibnamefont {Aharonovich}},\ }\bibfield  {title}
  {\bibinfo {title} {Quantum emission from hexagonal boron nitride
  monolayers},\ }\href {\doibase 10.1038/nnano.2015.242} {\bibfield  {journal}
  {\bibinfo  {journal} {Nat. Nanotech.}\ }\textbf {\bibinfo {volume} {11}},\
  \bibinfo {pages} {37} (\bibinfo {year} {2016})}\BibitemShut {NoStop}%
\bibitem [{\citenamefont {Park}\ \emph {et~al.}(2015)\citenamefont {Park},
  \citenamefont {Guo}, \citenamefont {Makarov},\ and\ \citenamefont
  {Klimov}}]{park:2015}%
  \BibitemOpen
  \bibfield  {author} {\bibinfo {author} {\bibfnamefont {Y.-S.}\ \bibnamefont
  {Park}}, \bibinfo {author} {\bibfnamefont {S.}~\bibnamefont {Guo}}, \bibinfo
  {author} {\bibfnamefont {N.~S.}\ \bibnamefont {Makarov}}, \ and\ \bibinfo
  {author} {\bibfnamefont {V.~I.}\ \bibnamefont {Klimov}},\ }\bibfield  {title}
  {\bibinfo {title} {Room temperature single-photon emission from individual
  perovskite quantum dots},\ }\href {\doibase 10.1021/acsnano.5b04584}
  {\bibfield  {journal} {\bibinfo  {journal} {ACS Nano}\ }\textbf {\bibinfo
  {volume} {9}},\ \bibinfo {pages} {10386} (\bibinfo {year}
  {2015})}\BibitemShut {NoStop}%
\bibitem [{\citenamefont {Wang}\ \emph {et~al.}(2005)\citenamefont {Wang},
  \citenamefont {Dukovic}, \citenamefont {Brus},\ and\ \citenamefont
  {Heinz}}]{Wang:2005}%
  \BibitemOpen
  \bibfield  {author} {\bibinfo {author} {\bibfnamefont {F.}~\bibnamefont
  {Wang}}, \bibinfo {author} {\bibfnamefont {G.}~\bibnamefont {Dukovic}},
  \bibinfo {author} {\bibfnamefont {L.~E.}\ \bibnamefont {Brus}}, \ and\
  \bibinfo {author} {\bibfnamefont {T.~F.}\ \bibnamefont {Heinz}},\ }\bibfield
  {title} {\bibinfo {title} {The optical resonances in carbon nanotubes arise
  from excitons},\ }\href {\doibase 10.1126/science.1110265} {\bibfield
  {journal} {\bibinfo  {journal} {Science}\ }\textbf {\bibinfo {volume}
  {308}},\ \bibinfo {pages} {838} (\bibinfo {year} {2005})}\BibitemShut
  {NoStop}%
\bibitem [{\citenamefont {Maultzsch}\ \emph {et~al.}(2005)\citenamefont
  {Maultzsch}, \citenamefont {Pomraenke}, \citenamefont {Reich}, \citenamefont
  {Chang}, \citenamefont {Prezzi}, \citenamefont {Ruini}, \citenamefont
  {Molinari}, \citenamefont {Strano}, \citenamefont {Thomsen},\ and\
  \citenamefont {Lienau}}]{Maultzsch:2005}%
  \BibitemOpen
  \bibfield  {author} {\bibinfo {author} {\bibfnamefont {J.}~\bibnamefont
  {Maultzsch}}, \bibinfo {author} {\bibfnamefont {R.}~\bibnamefont
  {Pomraenke}}, \bibinfo {author} {\bibfnamefont {S.}~\bibnamefont {Reich}},
  \bibinfo {author} {\bibfnamefont {E.}~\bibnamefont {Chang}}, \bibinfo
  {author} {\bibfnamefont {D.}~\bibnamefont {Prezzi}}, \bibinfo {author}
  {\bibfnamefont {A.}~\bibnamefont {Ruini}}, \bibinfo {author} {\bibfnamefont
  {E.}~\bibnamefont {Molinari}}, \bibinfo {author} {\bibfnamefont {M.~S.}\
  \bibnamefont {Strano}}, \bibinfo {author} {\bibfnamefont {C.}~\bibnamefont
  {Thomsen}}, \ and\ \bibinfo {author} {\bibfnamefont {C.}~\bibnamefont
  {Lienau}},\ }\bibfield  {title} {\bibinfo {title} {Exciton binding energies
  in carbon nanotubes from two-photon photoluminescence},\ }\href {\doibase 10.1103/PhysRevB.72.241402} {\bibfield  {journal} {\bibinfo  {journal} {Phys.
  Rev. B}\ }\textbf {\bibinfo {volume} {72}},\ \bibinfo {pages} {241402(R)}
  (\bibinfo {year} {2005})}\BibitemShut {NoStop}%
\bibitem [{\citenamefont {Weisman}\ and\ \citenamefont
  {Bachilo}(2003)}]{Weisman:2003}%
  \BibitemOpen
  \bibfield  {author} {\bibinfo {author} {\bibfnamefont {R.~B.}\ \bibnamefont
  {Weisman}}\ and\ \bibinfo {author} {\bibfnamefont {S.~M.}\ \bibnamefont
  {Bachilo}},\ }\bibfield  {title} {\bibinfo {title} {Dependence of optical
  transition energies on structure for single-walled carbon nanotubes in
  aqueous suspension: An empirical {K}ataura plot},\ }\href {\doibase 10.1021/nl034428i} {\bibfield  {journal} {\bibinfo  {journal} {Nano Lett.}\
  }\textbf {\bibinfo {volume} {3}},\ \bibinfo {pages} {1235} (\bibinfo {year}
  {2003})}\BibitemShut {NoStop}%
\bibitem [{\citenamefont {Ishii}\ \emph {et~al.}(2015)\citenamefont {Ishii},
  \citenamefont {Yoshida},\ and\ \citenamefont {Kato}}]{Ishii:2015}%
  \BibitemOpen
  \bibfield  {author} {\bibinfo {author} {\bibfnamefont {A.}~\bibnamefont
  {Ishii}}, \bibinfo {author} {\bibfnamefont {M.}~\bibnamefont {Yoshida}}, \
  and\ \bibinfo {author} {\bibfnamefont {Y.~K.}\ \bibnamefont {Kato}},\
  }\bibfield  {title} {\bibinfo {title} {Exciton diffusion, end quenching, and
  exciton-exciton annihilation in individual air-suspended carbon nanotubes},\
  }\href {\doibase 10.1103/PhysRevB.91.125427} {\bibfield  {journal} {\bibinfo
  {journal} {Phys. Rev. B}\ }\textbf {\bibinfo {volume} {91}},\ \bibinfo
  {pages} {125427} (\bibinfo {year} {2015})}\BibitemShut {NoStop}%
\bibitem [{\citenamefont {H\"ogele}\ \emph {et~al.}(2008)\citenamefont
  {H\"ogele}, \citenamefont {Galland}, \citenamefont {Winger},\ and\
  \citenamefont {Imamo\u{g}lu}}]{Hogele:2008}%
  \BibitemOpen
  \bibfield  {author} {\bibinfo {author} {\bibfnamefont {A.}~\bibnamefont
  {H\"ogele}}, \bibinfo {author} {\bibfnamefont {C.}~\bibnamefont {Galland}},
  \bibinfo {author} {\bibfnamefont {M.}~\bibnamefont {Winger}}, \ and\ \bibinfo
  {author} {\bibfnamefont {A.}~\bibnamefont {Imamo\u{g}lu}},\ }\bibfield
  {title} {\bibinfo {title} {Photon antibunching in the photoluminescence
  spectra of a single carbon nanotube},\ }\href {\doibase 10.1103/PhysRevLett.100.217401} {\bibfield  {journal} {\bibinfo  {journal}
  {Phys. Rev. Lett.}\ }\textbf {\bibinfo {volume} {100}},\ \bibinfo {pages}
  {217401} (\bibinfo {year} {2008})}\BibitemShut {NoStop}%
\bibitem [{\citenamefont {Hofmann}\ \emph {et~al.}(2013)\citenamefont
  {Hofmann}, \citenamefont {Gl\"uckert}, \citenamefont {No\'e}, \citenamefont
  {Bourjau}, \citenamefont {Dehmel},\ and\ \citenamefont
  {H\"ogele}}]{Hofmann:2013}%
  \BibitemOpen
  \bibfield  {author} {\bibinfo {author} {\bibfnamefont {M.~S.}\ \bibnamefont
  {Hofmann}}, \bibinfo {author} {\bibfnamefont {J.~T.}\ \bibnamefont
  {Gl\"uckert}}, \bibinfo {author} {\bibfnamefont {J.}~\bibnamefont {No\'e}},
  \bibinfo {author} {\bibfnamefont {C.}~\bibnamefont {Bourjau}}, \bibinfo
  {author} {\bibfnamefont {R.}~\bibnamefont {Dehmel}}, \ and\ \bibinfo {author}
  {\bibfnamefont {A.}~\bibnamefont {H\"ogele}},\ }\bibfield  {title} {\bibinfo
  {title} {Bright, long-lived and coherent excitons in carbon nanotube quantum
  dots},\ }\href {\doibase 10.1038/nnano.2013.119} {\bibfield  {journal}
  {\bibinfo  {journal} {Nat. Nanotech.}\ }\textbf {\bibinfo {volume} {8}},\
  \bibinfo {pages} {502} (\bibinfo {year} {2013})}\BibitemShut {NoStop}%
\bibitem [{\citenamefont {Walden-Newman}\ \emph {et~al.}(2012)\citenamefont
  {Walden-Newman}, \citenamefont {Sarpkaya},\ and\ \citenamefont
  {Strauf}}]{Newman:2012}%
  \BibitemOpen
  \bibfield  {author} {\bibinfo {author} {\bibfnamefont {W.}~\bibnamefont
  {Walden-Newman}}, \bibinfo {author} {\bibfnamefont {I.}~\bibnamefont
  {Sarpkaya}}, \ and\ \bibinfo {author} {\bibfnamefont {S.}~\bibnamefont
  {Strauf}},\ }\bibfield  {title} {\bibinfo {title} {Quantum light signatures
  and nanosecond spectral diffusion from cavity-embedded carbon nanotubes},\
  }\href {\doibase 10.1021/nl204402v} {\bibfield  {journal} {\bibinfo
  {journal} {Nano Lett.}\ }\textbf {\bibinfo {volume} {12}},\ \bibinfo {pages}
  {1934} (\bibinfo {year} {2012})}\BibitemShut {NoStop}%
\bibitem [{\citenamefont {Jeantet}\ \emph {et~al.}(2016)\citenamefont
  {Jeantet}, \citenamefont {Chassagneux}, \citenamefont {Raynaud},
  \citenamefont {Roussignol}, \citenamefont {Lauret}, \citenamefont {Besga},
  \citenamefont {Est\`eve}, \citenamefont {Reichel},\ and\ \citenamefont
  {Voisin}}]{Jeantet:2016}%
  \BibitemOpen
  \bibfield  {author} {\bibinfo {author} {\bibfnamefont {A.}~\bibnamefont
  {Jeantet}}, \bibinfo {author} {\bibfnamefont {Y.}~\bibnamefont
  {Chassagneux}}, \bibinfo {author} {\bibfnamefont {C.}~\bibnamefont
  {Raynaud}}, \bibinfo {author} {\bibfnamefont {Ph.}~\bibnamefont {Roussignol}},
  \bibinfo {author} {\bibfnamefont {J.~S.}\ \bibnamefont {Lauret}}, \bibinfo
  {author} {\bibfnamefont {B.}~\bibnamefont {Besga}}, \bibinfo {author}
  {\bibfnamefont {J.}~\bibnamefont {Est\`eve}}, \bibinfo {author}
  {\bibfnamefont {J.}~\bibnamefont {Reichel}}, \ and\ \bibinfo {author}
  {\bibfnamefont {C.}~\bibnamefont {Voisin}},\ }\bibfield  {title} {\bibinfo
  {title} {Widely tunable single-photon source from a carbon nanotube in the
  {P}urcell regime},\ }\href {\doibase 10.1103/PhysRevLett.116.247402}
  {\bibfield  {journal} {\bibinfo  {journal} {Phys. Rev. Lett.}\ }\textbf
  {\bibinfo {volume} {116}},\ \bibinfo {pages} {247402} (\bibinfo {year}
  {2016})}\BibitemShut {NoStop}%
\bibitem [{\citenamefont {Khasminskaya}\ \emph {et~al.}(2016)\citenamefont
  {Khasminskaya}, \citenamefont {Pyatkov}, \citenamefont {S\l{}owik},
  \citenamefont {Ferrari}, \citenamefont {Kahl}, \citenamefont {Kovalyuk},
  \citenamefont {Rath}, \citenamefont {Vetter}, \citenamefont {Hennrich},
  \citenamefont {Kappes}, \citenamefont {Gol'tsman}, \citenamefont {Korneev},
  \citenamefont {Rockstuhl}, \citenamefont {Krupke},\ and\ \citenamefont
  {Pernice}}]{Khasminskaya:2016}%
  \BibitemOpen
  \bibfield  {author} {\bibinfo {author} {\bibfnamefont {S.}~\bibnamefont
  {Khasminskaya}}, \bibinfo {author} {\bibfnamefont {F.}~\bibnamefont
  {Pyatkov}}, \bibinfo {author} {\bibfnamefont {K.}~\bibnamefont {S\l{}owik}},
  \bibinfo {author} {\bibfnamefont {S.}~\bibnamefont {Ferrari}}, \bibinfo
  {author} {\bibfnamefont {O.}~\bibnamefont {Kahl}}, \bibinfo {author}
  {\bibfnamefont {V.}~\bibnamefont {Kovalyuk}}, \bibinfo {author}
  {\bibfnamefont {P.}~\bibnamefont {Rath}}, \bibinfo {author} {\bibfnamefont
  {A.}~\bibnamefont {Vetter}}, \bibinfo {author} {\bibfnamefont
  {F.}~\bibnamefont {Hennrich}}, \bibinfo {author} {\bibfnamefont {M.~M.}\
  \bibnamefont {Kappes}}, \bibinfo {author} {\bibfnamefont {G.}~\bibnamefont
  {Gol'tsman}}, \bibinfo {author} {\bibfnamefont {A.}~\bibnamefont {Korneev}},
  \bibinfo {author} {\bibfnamefont {C.}~\bibnamefont {Rockstuhl}}, \bibinfo
  {author} {\bibfnamefont {R.}~\bibnamefont {Krupke}}, \ and\ \bibinfo {author}
  {\bibfnamefont {W.~H.~P.}\ \bibnamefont {Pernice}},\ }\bibfield  {title}
  {\bibinfo {title} {Fully integrated quantum photonic circuit with an
  electrically driven light source},\ }\href {\doibase 10.1038/nphoton.2016.178} {\bibfield  {journal} {\bibinfo  {journal} {Nat.
  Photon.}\ }\textbf {\bibinfo {volume} {10}},\ \bibinfo {pages} {727}
  (\bibinfo {year} {2016})}\BibitemShut {NoStop}%
\bibitem [{\citenamefont {Ma}\ \emph {et~al.}(2015{\natexlab{a}})\citenamefont
  {Ma}, \citenamefont {Hartmann}, \citenamefont {Baldwin}, \citenamefont
  {Doorn},\ and\ \citenamefont {Htoon}}]{Ma:2015NatNano}%
  \BibitemOpen
  \bibfield  {author} {\bibinfo {author} {\bibfnamefont {X.}~\bibnamefont
  {Ma}}, \bibinfo {author} {\bibfnamefont {N.~F.}\ \bibnamefont {Hartmann}},
  \bibinfo {author} {\bibfnamefont {J.~K.~S.}\ \bibnamefont {Baldwin}},
  \bibinfo {author} {\bibfnamefont {S.~K.}\ \bibnamefont {Doorn}}, \ and\
  \bibinfo {author} {\bibfnamefont {H.}~\bibnamefont {Htoon}},\ }\bibfield
  {title} {\bibinfo {title} {Room-temperature single-photon generation from
  solitary dopants of carbon nanotubes},\ }\href {\doibase 10.1038/nnano.2015.136} {\bibfield  {journal} {\bibinfo  {journal} {Nat.
  Nanotech.}\ }\textbf {\bibinfo {volume} {10}},\ \bibinfo {pages} {671}
  (\bibinfo {year} {2015}{\natexlab{a}})}\BibitemShut {NoStop}%
\bibitem [{\citenamefont {Endo}\ \emph {et~al.}(2015)\citenamefont {Endo},
  \citenamefont {Ishi-Hayase},\ and\ \citenamefont {Maki}}]{Endo:2015}%
  \BibitemOpen
  \bibfield  {author} {\bibinfo {author} {\bibfnamefont {T.}~\bibnamefont
  {Endo}}, \bibinfo {author} {\bibfnamefont {J.}~\bibnamefont {Ishi-Hayase}}, \
  and\ \bibinfo {author} {\bibfnamefont {H.}~\bibnamefont {Maki}},\ }\bibfield
  {title} {\bibinfo {title} {Photon antibunching in single-walled carbon
  nanotubes at telecommunication wavelengths and room temperature},\ }\href
  {\doibase 10.1063/1.4915618} {\bibfield  {journal} {\bibinfo  {journal}
  {Appl. Phys. Lett.}\ }\textbf {\bibinfo {volume} {106}},\ \bibinfo {pages}
  {113106} (\bibinfo {year} {2015})}\BibitemShut {NoStop}%
\bibitem [{\citenamefont {Wang}\ \emph {et~al.}(2004)\citenamefont {Wang},
  \citenamefont {Dukovic}, \citenamefont {Knoesel}, \citenamefont {Brus},\ and\
  \citenamefont {Heinz}}]{Wang:2004prb}%
  \BibitemOpen
  \bibfield  {author} {\bibinfo {author} {\bibfnamefont {F.}~\bibnamefont
  {Wang}}, \bibinfo {author} {\bibfnamefont {G.}~\bibnamefont {Dukovic}},
  \bibinfo {author} {\bibfnamefont {E.}~\bibnamefont {Knoesel}}, \bibinfo
  {author} {\bibfnamefont {L.~E.}\ \bibnamefont {Brus}}, \ and\ \bibinfo
  {author} {\bibfnamefont {T.~F.}\ \bibnamefont {Heinz}},\ }\bibfield  {title}
  {\bibinfo {title} {Observation of rapid {A}uger recombination in optically
  excited semiconducting carbon nanotubes},\ }\href {\doibase 10.1103/PhysRevB.70.241403} {\bibfield  {journal} {\bibinfo  {journal} {Phys.
  Rev. B}\ }\textbf {\bibinfo {volume} {70}},\ \bibinfo {pages} {241403(R)}
  (\bibinfo {year} {2004})}\BibitemShut {NoStop}%
\bibitem [{\citenamefont {Wang}\ \emph {et~al.}(2006)\citenamefont {Wang},
  \citenamefont {Wu}, \citenamefont {Hybertsen},\ and\ \citenamefont
  {Heinz}}]{Wang:2006}%
  \BibitemOpen
  \bibfield  {author} {\bibinfo {author} {\bibfnamefont {F.}~\bibnamefont
  {Wang}}, \bibinfo {author} {\bibfnamefont {Y.}~\bibnamefont {Wu}}, \bibinfo
  {author} {\bibfnamefont {M.~S.}\ \bibnamefont {Hybertsen}}, \ and\ \bibinfo
  {author} {\bibfnamefont {T.~F.}\ \bibnamefont {Heinz}},\ }\bibfield  {title}
  {\bibinfo {title} {Auger recombination of excitons in one-dimensional
  systems},\ }\href {\doibase 10.1103/PhysRevB.73.245424} {\bibfield  {journal}
  {\bibinfo  {journal} {Phys. Rev. B}\ }\textbf {\bibinfo {volume} {73}},\
  \bibinfo {pages} {245424} (\bibinfo {year} {2006})}\BibitemShut {NoStop}%
\bibitem [{\citenamefont {de~Gennes}(1982)}]{Gennes:1982}%
  \BibitemOpen
  \bibfield  {author} {\bibinfo {author} {\bibfnamefont {P.~G.}\ \bibnamefont
  {de~Gennes}},\ }\bibfield  {title} {\bibinfo {title} {Kinetics of
  diffusion-controlled processes in dense polymer systems. {I}. {N}onentangled
  regimes},\ }\href {\doibase 10.1063/1.443328} {\bibfield  {journal} {\bibinfo
   {journal} {J. Chem. Phys.}\ }\textbf {\bibinfo {volume} {76}},\ \bibinfo
  {pages} {3316} (\bibinfo {year} {1982})}\BibitemShut {NoStop}%
\bibitem [{\citenamefont {Ma}\ \emph {et~al.}(2015{\natexlab{b}})\citenamefont
  {Ma}, \citenamefont {Roslyak}, \citenamefont {Duque}, \citenamefont {Pang},
  \citenamefont {Doorn}, \citenamefont {Piryatinski}, \citenamefont {Dunlap},\
  and\ \citenamefont {Htoon}}]{Ma:2015PRL}%
  \BibitemOpen
  \bibfield  {author} {\bibinfo {author} {\bibfnamefont {X.}~\bibnamefont
  {Ma}}, \bibinfo {author} {\bibfnamefont {O.}~\bibnamefont {Roslyak}},
  \bibinfo {author} {\bibfnamefont {J.~G.}\ \bibnamefont {Duque}}, \bibinfo
  {author} {\bibfnamefont {X.}~\bibnamefont {Pang}}, \bibinfo {author}
  {\bibfnamefont {S.~K.}\ \bibnamefont {Doorn}}, \bibinfo {author}
  {\bibfnamefont {A.}~\bibnamefont {Piryatinski}}, \bibinfo {author}
  {\bibfnamefont {D.~H.}\ \bibnamefont {Dunlap}}, \ and\ \bibinfo {author}
  {\bibfnamefont {H.}~\bibnamefont {Htoon}},\ }\bibfield  {title} {\bibinfo
  {title} {Influences of exciton diffusion and exciton-exciton annihilation on
  photon emission statistics of carbon nanotubes},\ }\href {\doibase 10.1103/PhysRevLett.115.017401} {\bibfield  {journal} {\bibinfo  {journal}
  {Phys. Rev. Lett.}\ }\textbf {\bibinfo {volume} {115}},\ \bibinfo {pages}
  {017401} (\bibinfo {year} {2015}{\natexlab{b}})}\BibitemShut {NoStop}%
\bibitem [{\citenamefont {Yoshikawa}\ \emph {et~al.}(2010)\citenamefont
  {Yoshikawa}, \citenamefont {Matsuda},\ and\ \citenamefont
  {Kanemitsu}}]{Yoshikawa:2010}%
  \BibitemOpen
  \bibfield  {author} {\bibinfo {author} {\bibfnamefont {K.}~\bibnamefont
  {Yoshikawa}}, \bibinfo {author} {\bibfnamefont {K.}~\bibnamefont {Matsuda}},
  \ and\ \bibinfo {author} {\bibfnamefont {Y.}~\bibnamefont {Kanemitsu}},\
  }\bibfield  {title} {\bibinfo {title} {Exciton transport in suspended single
  carbon nanotubes studied by photoluminescence imaging spectroscopy},\ }\href
  {\doibase 10.1021/jp911518h} {\bibfield  {journal} {\bibinfo  {journal} {J.
  Phys. Chem. C}\ }\textbf {\bibinfo {volume} {114}},\ \bibinfo {pages} {4353}
  (\bibinfo {year} {2010})}\BibitemShut {NoStop}%
\bibitem [{\citenamefont {Harrah}\ and\ \citenamefont
  {Swan}(2011)}]{Harrah:2011a}%
  \BibitemOpen
  \bibfield  {author} {\bibinfo {author} {\bibfnamefont {D.~M.}\ \bibnamefont
  {Harrah}}\ and\ \bibinfo {author} {\bibfnamefont {A.~K.}\ \bibnamefont
  {Swan}},\ }\bibfield  {title} {\bibinfo {title} {The role of length and
  defects on optical quantum efficiency and exciton decay dynamics in
  single-walled carbon nanotubes},\ }\href {\doibase 10.1021/nn1031214}
  {\bibfield  {journal} {\bibinfo  {journal} {ACS Nano}\ }\textbf {\bibinfo
  {volume} {5}},\ \bibinfo {pages} {647} (\bibinfo {year} {2011})}\BibitemShut
  {NoStop}%
\bibitem [{\citenamefont {Moritsubo}\ \emph {et~al.}(2010)\citenamefont
  {Moritsubo}, \citenamefont {Murai}, \citenamefont {Shimada}, \citenamefont
  {Murakami}, \citenamefont {Chiashi}, \citenamefont {Maruyama},\ and\
  \citenamefont {Kato}}]{Moritsubo:2010}%
  \BibitemOpen
  \bibfield  {author} {\bibinfo {author} {\bibfnamefont {S.}~\bibnamefont
  {Moritsubo}}, \bibinfo {author} {\bibfnamefont {T.}~\bibnamefont {Murai}},
  \bibinfo {author} {\bibfnamefont {T.}~\bibnamefont {Shimada}}, \bibinfo
  {author} {\bibfnamefont {Y.}~\bibnamefont {Murakami}}, \bibinfo {author}
  {\bibfnamefont {S.}~\bibnamefont {Chiashi}}, \bibinfo {author} {\bibfnamefont
  {S.}~\bibnamefont {Maruyama}}, \ and\ \bibinfo {author} {\bibfnamefont
  {Y.~K.}\ \bibnamefont {Kato}},\ }\bibfield  {title} {\bibinfo {title}
  {Exciton diffusion in air-suspended single-walled carbon nanotubes},\ }\href
  {\doibase 10.1103/PhysRevLett.104.247402} {\bibfield  {journal} {\bibinfo
  {journal} {Phys. Rev. Lett.}\ }\textbf {\bibinfo {volume} {104}},\ \bibinfo
  {pages} {247402} (\bibinfo {year} {2010})}\BibitemShut {NoStop}%
\bibitem [{\citenamefont {Murakami}\ and\ \citenamefont
  {Kono}(2009)}]{Murakami:2009prb2}%
  \BibitemOpen
  \bibfield  {author} {\bibinfo {author} {\bibfnamefont {Y.}~\bibnamefont
  {Murakami}}\ and\ \bibinfo {author} {\bibfnamefont {J.}~\bibnamefont
  {Kono}},\ }\bibfield  {title} {\bibinfo {title} {Existence of an upper limit
  on the density of excitons in carbon nanotubes by diffusion-limited
  exciton-exciton annihilation: Experiment and theory},\ }\href {\doibase 10.1103/PhysRevB.80.035432} {\bibfield  {journal} {\bibinfo  {journal} {Phys.
  Rev. B}\ }\textbf {\bibinfo {volume} {80}},\ \bibinfo {pages} {035432}
  (\bibinfo {year} {2009})}\BibitemShut {NoStop}%
\bibitem [{\citenamefont {Xiao}\ \emph {et~al.}(2010)\citenamefont {Xiao},
  \citenamefont {Nhan}, \citenamefont {Wilson},\ and\ \citenamefont
  {Fraser}}]{Xiao:2010}%
  \BibitemOpen
  \bibfield  {author} {\bibinfo {author} {\bibfnamefont {Y.-F.}\ \bibnamefont
  {Xiao}}, \bibinfo {author} {\bibfnamefont {T.~Q.}\ \bibnamefont {Nhan}},
  \bibinfo {author} {\bibfnamefont {M.~W.~B.}\ \bibnamefont {Wilson}}, \ and\
  \bibinfo {author} {\bibfnamefont {J.~M.}\ \bibnamefont {Fraser}},\ }\bibfield
   {title} {\bibinfo {title} {Saturation of the photoluminescence at
  few-exciton levels in a single-walled carbon nanotube under ultrafast
  excitation},\ }\href {\doibase 10.1103/PhysRevLett.104.017401} {\bibfield
  {journal} {\bibinfo  {journal} {Phys. Rev. Lett.}\ }\textbf {\bibinfo
  {volume} {104}},\ \bibinfo {pages} {017401} (\bibinfo {year}
  {2010})}\BibitemShut {NoStop}%
\bibitem [{\citenamefont {Nair}\ \emph {et~al.}(2011)\citenamefont {Nair},
  \citenamefont {Zhao},\ and\ \citenamefont {Bawendi}}]{Nair:2011}%
  \BibitemOpen
  \bibfield  {author} {\bibinfo {author} {\bibfnamefont {G.}~\bibnamefont
  {Nair}}, \bibinfo {author} {\bibfnamefont {J.}~\bibnamefont {Zhao}}, \ and\
  \bibinfo {author} {\bibfnamefont {M.~G.}\ \bibnamefont {Bawendi}},\
  }\bibfield  {title} {\bibinfo {title} {Biexciton quantum yield of single
  semiconductor nanocrystals from photon statistics},\ }\href {\doibase 10.1021/nl104054t} {\bibfield  {journal} {\bibinfo  {journal} {Nano Lett.}\
  }\textbf {\bibinfo {volume} {11}},\ \bibinfo {pages} {1136} (\bibinfo {year}
  {2011})}\BibitemShut {NoStop}%
\bibitem [{\citenamefont {Redner}(2001)}]{Redner}%
  \BibitemOpen
  \bibfield  {author} {\bibinfo {author} {\bibfnamefont {S.}~\bibnamefont
  {Redner}},\ }\href@noop {} {\emph {\bibinfo {title} {A Guide to First-Passage
  Processes}}}\ (\bibinfo  {publisher} {Cambridge Univ. Press},\ \bibinfo
  {year} {2001})\BibitemShut {NoStop}%
\bibitem [{\citenamefont {Rauhut}\ \emph {et~al.}(2012)\citenamefont {Rauhut},
  \citenamefont {Engel}, \citenamefont {Steiner}, \citenamefont {Krupke},
  \citenamefont {Avouris},\ and\ \citenamefont {Hartschuh}}]{Rauhut:2012}%
  \BibitemOpen
  \bibfield  {author} {\bibinfo {author} {\bibfnamefont {N.}~\bibnamefont
  {Rauhut}}, \bibinfo {author} {\bibfnamefont {M.}~\bibnamefont {Engel}},
  \bibinfo {author} {\bibfnamefont {M.}~\bibnamefont {Steiner}}, \bibinfo
  {author} {\bibfnamefont {R.}~\bibnamefont {Krupke}}, \bibinfo {author}
  {\bibfnamefont {P.}~\bibnamefont {Avouris}}, \ and\ \bibinfo {author}
  {\bibfnamefont {A.}~\bibnamefont {Hartschuh}},\ }\bibfield  {title} {\bibinfo
  {title} {Antenna-enhanced photocurrent microscopy on single-walled carbon
  nanotubes at 30 nm resolution},\ }\href {\doibase 10.1021/nn301979c}
  {\bibfield  {journal} {\bibinfo  {journal} {ACS Nano}\ }\textbf {\bibinfo
  {volume} {6}},\ \bibinfo {pages} {6416} (\bibinfo {year} {2012})}\BibitemShut
  {NoStop}%
\bibitem [{\citenamefont {Liao}\ \emph {et~al.}(2016)\citenamefont {Liao},
  \citenamefont {Jiang}, \citenamefont {Hu}, \citenamefont {Zhang},
  \citenamefont {Kuang}, \citenamefont {Zhu}, \citenamefont {Zhang},\ and\
  \citenamefont {Dong}}]{Liao:2016}%
  \BibitemOpen
  \bibfield  {author} {\bibinfo {author} {\bibfnamefont {M.}~\bibnamefont
  {Liao}}, \bibinfo {author} {\bibfnamefont {S.}~\bibnamefont {Jiang}},
  \bibinfo {author} {\bibfnamefont {C.}~\bibnamefont {Hu}}, \bibinfo {author}
  {\bibfnamefont {R.}~\bibnamefont {Zhang}}, \bibinfo {author} {\bibfnamefont
  {Y.}~\bibnamefont {Kuang}}, \bibinfo {author} {\bibfnamefont
  {J.}~\bibnamefont {Zhu}}, \bibinfo {author} {\bibfnamefont {Y.}~\bibnamefont
  {Zhang}}, \ and\ \bibinfo {author} {\bibfnamefont {Z.}~\bibnamefont {Dong}},\
  }\bibfield  {title} {\bibinfo {title} {Tip-enhanced {R}aman spectroscopic
  imaging of individual carbon nanotubes with subnanometer resolution},\ }\href
  {\doibase 10.1021/acs.nanolett.6b00533} {\bibfield  {journal} {\bibinfo
  {journal} {Nano Lett.}\ }\textbf {\bibinfo {volume} {16}},\ \bibinfo {pages}
  {4040} (\bibinfo {year} {2016})}\BibitemShut {NoStop}%
\bibitem [{\citenamefont {Tinnefeld}\ \emph {et~al.}(2001)\citenamefont
  {Tinnefeld}, \citenamefont {M\"{u}ller},\ and\ \citenamefont
  {Sauer}}]{Tinnefeld:2001}%
  \BibitemOpen
  \bibfield  {author} {\bibinfo {author} {\bibfnamefont {P.}~\bibnamefont
  {Tinnefeld}}, \bibinfo {author} {\bibfnamefont {C.}~\bibnamefont
  {M\"{u}ller}}, \ and\ \bibinfo {author} {\bibfnamefont {M.}~\bibnamefont
  {Sauer}},\ }\bibfield  {title} {\bibinfo {title} {Time-varying photon
  probability distribution of individual molecules at room temperature},\
  }\href {\doibase 10.1016/S0009-2614(01)00883-1} {\bibfield  {journal}
  {\bibinfo  {journal} {Chem. Phys. Lett.}\ }\textbf {\bibinfo {volume}
  {345}},\ \bibinfo {pages} {252 } (\bibinfo {year} {2001})}\BibitemShut
  {NoStop}%
\bibitem [{\citenamefont {Higashide}\ \emph {et~al.}(2017)\citenamefont
  {Higashide}, \citenamefont {Yoshida}, \citenamefont {Uda}, \citenamefont
  {Ishii},\ and\ \citenamefont {Kato}}]{Higashide:2017}%
  \BibitemOpen
  \bibfield  {author} {\bibinfo {author} {\bibfnamefont {N.}~\bibnamefont
  {Higashide}}, \bibinfo {author} {\bibfnamefont {M.}~\bibnamefont {Yoshida}},
  \bibinfo {author} {\bibfnamefont {T.}~\bibnamefont {Uda}}, \bibinfo {author}
  {\bibfnamefont {A.}~\bibnamefont {Ishii}}, \ and\ \bibinfo {author}
  {\bibfnamefont {Y.~K.}\ \bibnamefont {Kato}},\ }\bibfield  {title} {\bibinfo
  {title} {Cold exciton electroluminescence from air-suspended carbon nanotube
  split-gate devices},\ }\href {\doibase 10.1063/1.4983278} {\bibfield
  {journal} {\bibinfo  {journal} {Appl. Phys. Lett.}\ }\textbf {\bibinfo
  {volume} {110}},\ \bibinfo {pages} {191101} (\bibinfo {year}
  {2017})}\BibitemShut {NoStop}%
\bibitem [{\citenamefont {Miura}\ \emph {et~al.}(2014)\citenamefont {Miura},
  \citenamefont {Imamura}, \citenamefont {Ohta}, \citenamefont {Ishii},
  \citenamefont {Liu}, \citenamefont {Shimada}, \citenamefont {Iwamoto},
  \citenamefont {Arakawa},\ and\ \citenamefont {Kato}}]{Miura:2014}%
  \BibitemOpen
  \bibfield  {author} {\bibinfo {author} {\bibfnamefont {R.}~\bibnamefont
  {Miura}}, \bibinfo {author} {\bibfnamefont {S.}~\bibnamefont {Imamura}},
  \bibinfo {author} {\bibfnamefont {R.}~\bibnamefont {Ohta}}, \bibinfo {author}
  {\bibfnamefont {A.}~\bibnamefont {Ishii}}, \bibinfo {author} {\bibfnamefont
  {X.}~\bibnamefont {Liu}}, \bibinfo {author} {\bibfnamefont {T.}~\bibnamefont
  {Shimada}}, \bibinfo {author} {\bibfnamefont {S.}~\bibnamefont {Iwamoto}},
  \bibinfo {author} {\bibfnamefont {Y.}~\bibnamefont {Arakawa}}, \ and\
  \bibinfo {author} {\bibfnamefont {Y.~K.}\ \bibnamefont {Kato}},\ }\bibfield
  {title} {\bibinfo {title} {Ultralow mode-volume photonic crystal nanobeam
  cavities for high-efficiency coupling to individual carbon nanotube
  emitters},\ }\href {\doibase 10.1038/ncomms6580} {\bibfield  {journal}
  {\bibinfo  {journal} {Nat. Commun.}\ }\textbf {\bibinfo {volume} {5}},\
  \bibinfo {pages} {5580} (\bibinfo {year} {2014})}\BibitemShut {NoStop}%
\end{thebibliography}
\end{document}